\documentstyle[aaspp]{article}

\newlength{\parindnt}
\newcommand{\PSbox}[3]{\mbox{\rule{0in}{#3}\includegraphics{#1}\hspace{#2}}}
\newcommand{\FigNum}[1]{\unitlength 1pt \begin{picture}(55,10)(-400,35) 
			\put(0,0){Figure #1}
			\end{picture}}

\newcommand\B{{$B$}}
\newcommand\V{{$V$}}

\renewcommand\approx{\mbox{$\sim$}}
\newcommand\about{\mbox{$\sim$}}

\newcommand\approxgt{\mbox{$^{>}\hspace{-0.24cm}_{\sim}$}}
\newcommand\approxlt{\mbox{$^{<}\hspace{-0.24cm}_{\sim}$}}

\newcommand\ie{{\it i.e.}}
\newcommand\eg{{\it e.g.}}
\newcommand\cf{{\it cf.}}
\newcommand\etal{et~al.$\!$}

\def\Ref #1 {\lbrack {#1}\rbrack}
\def\vol#1  {{{\bf #1}{\rm,}\ }}
\def\etal   {{et~al.}\ }
\def\apj    {{ApJ{\rm,}\ }}
\def\apjl   {{ApJLett{\rm,}\ }}
\def\apjlett{{ApJLett{\rm,}\ }}
\def\apjs   {{ApJS{\rm,}\ }}

\def\aa     {{A\&A{\rm,}\ }}

\def\mnras  {{MNRAS{\rm,}\ }}
\def\nat    {{Nat{\rm,}\ }}

\begin{document}
\def\cgro{{\it C}GRO}
\def\t90{{t$_{90}$}}
\def\V{{\it V}}
\def\B{{\it B}}
\def\Fmean{{$<$F$>$}}
\def\bog{Paczy\'{n}ski}
\def\wick{{Wickramasinghe}}
\def\persec{{sec$^{-1}$}}
\def\percm{{cm$^{-2}$}}
\def\vvmax{{$<V/V_{\rm max}>$}}
\def\zmax{{$z_{\rm max}$}}
\def\zmin{{$z_{\rm min}$}}
\def\tdur{{$t_{\rm dur}$}}
\def\oo{{$\Omega_0$}}
\def\lnls{{log N -- log F$_{\rm peak}$}}
\def\lgs{{LGS}}
\def\lstar{{$L^*$}}
\def\nstar{{$n_*$}}
\def\nnot{{$n_0$}}
\def\plusz{{$(1+z)$}}
\def\rrel{{$R_{\rm rel}$}}
\def\falpha{{$f_{\alpha}$}}
\def\ebreak{$E_{\rm b}$}
\def\zbreak{{$z_{\rm b}$}}
\received{July 23, 1994}
\journalid{00}{MNRAS}
\begin{center}
Monthly Noticies of the Royal Astronomical Society, in press \\
\vspace{ 2cm}
{\bf Standard Cosmology and the BATSE Number vs Peak Flux Distribution} \\
{\sc Robert E. Rutledge, Lam Hui,  \& Walter H.~G. Lewin} \\
{\rm \small Dept. of Physics,  37-624B MIT, Cambridge, MA 02139 \\
 rutledge@space.mit.edu, lhui@space.mit.edu, \& lewin@space.mit.edu}
\end{center}

\begin{abstract}
Adopting a simple cosmological model for Gamma Ray Bursts (GRBs),
following Mao \& \bog~ (1992), we generate number vs. peak flux
distributions for a range of values of \oo~(ratio of the density of
the universe to the critical density) and \zmax~(the redshift at which
the faintest GRB in the present sample is located), and compare these
distributions to one from BATSE GRBs in the 2B catalog.  The observed
BATSE distribution is consistent with the faintest GRBs in our sample
originating from a redshift of \zmax~ $\approx$ 0.8-3.0 (90\%), with
the most likely values in the range of 1.0-2.2, and is largely
insensitive to \oo~ for models with no evolution.

To constrain the model parameter \oo~to the range 0.1-1.0 using only
\lnls~distributions, more than 4000 GRBs, with a most likely value of
$\sim9,000$ GRBs, above the 1024 msec averaged peak flux of 0.3 phot
\percm \persec would be needed.  This requires a live integration time
of $>6$ years with BATSE.  Detectors sensitive to much lower limits
(\ie~ standard candle bursts out to \zmax=10, $\sim \times$70-400 in
sensitivity) require $\approx$200 GRBs, with $<0.03$ year $4\pi$ ster
coverage.

We place limits on the amount of frequency density and, separately,
peak luminosity evolution in the sample of GRBs.  We find that
frequency density evolution models can place the faintest GRBs at
\zmax~\about 10-200, without conflicting with the observations of
relative time dilation of \about 2 reported by Norris \etal~(1994a)
and Wijers \& \bog\ (1994) (however, see \cite{band94}), although this
would require vastly different comoving burst rates in GRBs of
different spectra.  
 
\end{abstract}

\section{Introduction}

More than 20 years after their discovery (Klebesadel, Strong, \& Olson
1973), Gamma Ray Bursts (GRBs) continue to be elusive about their
origin or, possibly, origins.

A number of workers report results which are consistent with the
cosmological hypothesis, using data obtained by the Burst and
Transient Source Experiment (BATSE; \cite{fishman89} ) on the {\it
Compton} Gamma Ray Observatory (CGRO).  

A small subset of 'simple' single-peaked GRBs observed by BATSE showed
that bursts with the lowest peak fluxes had longer rise times than
GRBs with the highest peak fluxes (\cite{chryssa91}).

Dermer (1992) found that a cosmological model with no source source
evolution and a $q_0$=1/2 Freidmann cosmology could reproduce the
differential \vvmax distribution from 126 BATSE GRBs.  It was also
found that the observed \vvmax could accommodate moderate frequency
density evolution, with $(1+z)^p$, for $p=-2$, 0, and 2.

Piran (1992) found the BATSE-observed $<V/V_{\rm max}>$ statistic
(see, \eg, \cite{schmidt88}) and number distribution N($V/V_{\rm
max}$) are consistent with two flat-space cosmological models, one
with a zero cosmological constant and mass density $\Omega_{\rm M}$=1,
and a second with $\Omega_{\rm M}$=0.1 and a non-zero cosmological
constant ($\Lambda$), with the faintest GRBs coming from
\zmax \approx 1.  It was also found that the \vvmax distribution could
accommodate some frequency density evolution, with $(1+z)^p$, for
$p=-3$ to 3/2.

Mao \& \bog~(1992; hereafter MP), assuming no evolution in the burst
rate (a constant frequency density), and an
\oo=1,  $\Lambda=0$ universe, found that BATSE GRBs with assumed
photon number spectral indices in the range 0.5--1.5 (somewhat flatter
than actually observed by BATSE) would match the $<V/V_{\rm max}>$
value from the first $\sim$ 150 GRBs observed by BATSE
(\cite{meegan92}) if the faintest GRBs originated from
\zmax=1.1-2.6.

Lestrade \etal (1993), following a suggestion by \bog~(1992), used 20
GRBs observed by the PHEBUS instrument on board GRANAT with durations
$<$ 1.5 sec.  They showed that stronger bursts (in peak counts above
background in a 1/4 sec time bin) had a significantly shorter average
duration than weaker bursts. Using a Kolgomorov-Smirnov (KS) test, a
97\% probability that the stronger GRBs were drawn from a different
duration population than the weaker GRBs in this sample was found.

\wick\ \etal (1993) fit the \lnls\ curve of
118 BATSE GRBs with a model of GRBs of cosmological origin, assuming a
constant number of burst sources per comoving volume -- a different
assumption than that made by MP --  which does not take into account the
effect of time-dilation on the burst rate.  It was found that GRBs at
the completeness flux limit could originate at \zmax=0.4-1.7,
depending on the assumed source spectrum (photon number spectrum in
the range of 1.5-2.5; softer spectrum produced higher \zmax~ values).

Davis \etal (1993), using an analysis of pulse-width distributions
from 135 BATSE GRB lightcurves found that dim GRBs have pulse-widths
that are of $\sim$ 2 greater than that of bright bursts, consistent
with the dimmest GRBs coming from a z$\sim$1.

Tamblyn \& Melia (1993) found that, assuming a ``standard''
energy-spectral break at 300 keV, the average value of ratio of the
flux detection limit to the peak burst flux to the 3/2 power
($<(F_{min}/F_{max})^{3/2}>$) as a function of the observed full-sky
burst rate for several different GRB instruments (\ie\ PVO, SMM, KONUS,
APEX, SIGNE, and BATSE) is consistent with a non-evolving cosmological
population. 

Fenimore \etal (1993), using a composite intensity distribution of
BATSE and Pioneer Venus Orbiter (PVO) experiments, fit a homogeneous
model of standard candle sources, with the faintest BATSE bursts
coming from $z\sim0.80$; using model-derived distances to the
brightest sources within the error boxes of 8 bright GRBs, they found
that host galaxies must be fainter than an absolute magnitude of $-18$.

Lamb, Graziani \& Smith (1993) found that bursts which are ``smooth''
on short (\approx 64 ms) timescales have peak counts integrated on a
1024msec timescale (\B) which indicate ``smooth'' bursts are both
faint (\ie\ low number of counts) and bright (\ie\ high number of
counts), while bursts which are ``variable'' on short timescales are
faint only.  The ``smooth'' and ``variable'' classification is based
on a parameter $V$, the ratio of the peak counts in a 64 msec time bin
to the peak counts in a 1024 msec time bin ($C^{64}_{\rm
max}/C^{1024}_{\rm max}$), measured by the second most brightly
illuminated Large Area Detector of BATSE.  There is a known systematic
correlation of \V~with burst duration (\cite{rutledge93}); the minimum
\V~value is a function of duration $t_{\rm dur}$, the result
of which is that all GRBs of \tdur $\approxlt 400$ msec are included
in the ``variable'' class.  These short GRBs make up $\sim$ 2/3 of the
``variable'' class.

It might thus appear that the conclusion of
\lgs~requires that short GRBs be faint, while long GRBs be both bright
and faint, in apparent contradiction to the conclusions of Lestrade
\etal (1993).  However, since the durations of these short,
``variable'' bursts are less than the sampling period, \B~ represents
a measure of counts-fluence, rather than a sampled instantaneous peak flux.
Thus, one should not conclude that the results of \lgs~ are in
contradiction to those of Lestrade \etal (1993), (specifically, that
shorter GRBs have higher peak fluxes than longer GRBs, consistent with
the time dilation effect predicted by \bog~1992).

Norris \etal (1994a), using three tests to compare the time-scales of
131 bright and dim GRBs observed by BATSE of duration greater than 1.5
sec, finds temporal, intensity, and fluence scaling effects consistent
with a cosmological time dilation, with results consistent with the
dimmest GRBs coming from a z$\sim$1.

Wijers \& \bog~ (1994) analyzed the time dilations of bright and faint
samples of bursts, and found the relative duration distribution widths
were the same.  They concluded that the effect is more convincingly
explained by a cosmological origin than by a distribution intrinsic to
the source population.

A cosmological source of GRBs is not necessarily the sole explanation of the
results of these analyses. It cannot be ruled out that the results are
consistent with the intrinsic characteristics of a population located
in an extended galactic halo (or, corona).  Hakkila \etal~ (1994)
recently produced a detailed study of the constraints on galactic
populations from BATSE observations.

Here, we use the cosmological model for GRBs suggested by MP, except
we adjust \oo~and \zmax~to find \lnls~distributions consistent with
BATSE observations, for a range of assumed energy spectra, in order to
set limits on the acceptable parameter space for non-evolving
cosmological models, and for cosmological models with evolving
frequency density and luminosity.  We calculate the number of GRBs
which would be required to differentiate between values of \oo~of 0.1
and 1, and also between values of 0.1 and 0.5, depending on the
sensitivity of the detector as indicated by \zmax.  Finally we
calculate the amount of integration time required to distinguish
between these two values of \oo.

\section{Analysis}

Throughout this paper, we assume a cosmological constant $\Lambda$=0,
and a universe dominated by non-relativistic matter, which sets
$\Omega_0=2 q_0$, and spherical symmetry of radiation (no beaming).
The effects of beaming on GRB observations have been considered
elsewhere (Mao \& Yi 1994). 

We also make extensive use of the Kolgomorov-Smirnov test for
comparing an observed distribution with a model distribution (\cf\
Press \etal 1988, p. 491).  This test results in a probability that,
given the model distribution, another observed distribution as
disparate or more disparate could be drawn from the source population.
Thus, a KS probability of 1\% implies that only 1\% of the data sets
of the same size as the observed data set would be as or more
disparate from the model as the observed data set.  Subsequently, when
we say ``90\% confidence limits'', we mean the boundary around those
models with KS probabilities $>$10\%. 

\subsection{Number- Peak Flux distribution}
This discussion follows closely the model described by MP; we have
assumed their ``simplest'' model, with no adjustable parameters,
except we adjust one of the parameters (\oo).

\subsubsection{The Assumed Spectrum}
We here discuss our choice of the assumed GRB spectrum, as this
effects our results quantitatively, although not qualitatively, and
warrants clarification. 

We first assume that all GRBs have identical peak luminosities
(\ie~are standard candles), with (first) identical power law photon
number spectra:
\begin{eqnarray}
\label{eq:spectra}
\frac{dI_\nu}{d\nu}= C_z \nu^{-\alpha}  \\
C_z=C_0 (1 + z)^l \label{eq:specev}
\end{eqnarray}
where $I_\nu$ is in the units photons sec$^{-1}$, and $C_z$ is a
z-dependent peak luminosity normalization, which accounts for the peak
luminosity evolution with redshift.  We chose to represent the
evolution by power-law for its simplicity, unmotivated by any
theoretical expectation that it should actually follow a power-law
form. 

The exponent $\alpha$ has been found to be largely in the range of
1.5--2.5 (\cite{schaefer94}) for 30 GRBs selected from among 260 GRBs
in the first BATSE catalog (\cite{fishman94}).  We are motivated to
use a simple power-law spectrum, as a simple power-law spectrum (of
$\alpha$=2.0) was assumed in producing the observed peak photon flux
values in the BATSE 1B catalog.  Schaefer \etal (1994) found that
single power-law spectra adequately describe the observed GRB spectra
for the majority of bursts.  Using data from Table 5 Schaefer \etal
(1994) we produced the number histogram ( Fig~\ref{fig:alphahist}) of
the 30 measured $\alpha$, for fits over all ``valid energy ranges''.
The majority of these GRBs have $\alpha$ between 1.5 and 2.5. In fits
over energy ranges $<$ 500 keV, the resultant values of $\alpha$ tend
to be harder.  It should be noted that the GRBs in this sample were
selected for their high statistics, and it is known that GRBs which
have higher peak fluxes are spectrally harder than bursts with lower
peak fluxes (\cite{norris94}).

The single power-law energy spectrum is an over-simplification of the
true GRB energy spectrum.   Band
\etal (1993) found that GRB energy spectra are well described at low
energies by a power-law with an exponential cutoff, and at high
energies by a simple (steeper) power-law, of the form:
\begin{eqnarray}
\frac{dI_\nu}{d\nu} &= C_z \nu^{-\alpha_b} \exp ^{-\nu/\nu_{\rm b}}       &  \mbox {$\nu < \nu_{\rm b} ( \alpha_b - \beta )$} \label{eq:bandspectra1} \\
                    &= C_z \nu_{\rm b}^{-\alpha_b} \exp^{-1} \nu^{-\beta}  &  \mbox {$\nu >  \nu_{\rm b} ( \alpha_b - \beta )$ \label{eq:bandspectra2}}  
\end{eqnarray}
\noindent where $C_z$ is as in Eq.~\ref{eq:specev} (here, we use
$\alpha_b$ to differentiate between this power law and the single
power-law parameter $\alpha$).  In Fig.~\ref{fig:bandspec}a, we show
the fit spectral models for the $\approx$ 55 bursts produced by Band
\etal (1993) (their Table 4), normalized at 50.0 keV, over the energy range
50-5000 keV.  As Band \etal concluded, there is no ``universal''
spectrum, and this complicates interpretation of analyses which rely on
a single assumed ``universal'' spectrum, such as has been performed by
Tamblyn and Melia (1993).  However, a spectrum which largely follows
the behavior of the burst spectra, has the values $\alpha_b$=1.0,
$\beta$ = 2.5, and $h\nu_{\rm b}$ = 300 keV (hereafter, the complex
spectrum).  Further, the majority of these spectral models are steeper
than a single power-law $\alpha=1.0$, but more shallow than a single
power-law $\alpha=2.5$.  We show these three models in
Fig.~\ref{fig:bandspec}b; the complex spectrum lies at all times
between the single power-law spectra $\alpha=1.0$ and 2.5.

Our approach here is to perform the analyses separately for two
single-power law spectral slopes ($\alpha$=1.0, and 2.5) and for
the above-defined complex spectrum.  Thus, the acceptable parameter
space of cosmological models will be conservatively defined by the
acceptable parameter space spanned by models for each of the assumed
energy spectra.  Our use of single power-law spectra will be justified
by the qualitatively similar behavior between model parameter space
using the complex spectra and that using single power-law spectra. 

\subsubsection{Procedure for Single Power-law Spectra}
We assume the burst detectors are sensitive only to photons within a
fixed frequency range \label{sec:passband} ($\nu_{1} \le \nu \le
\nu_{2}$).  For a source with a single power-law photon spectrum, located at
redshift z, the part of the spectrum which is shifted into the
detector bandwidth is:
\begin{eqnarray}
\label{eq:peaklum}
I(z) = \int_{\nu_1(1+z)}^{\nu_2(1+z)} \frac{dI_\nu}{d\nu} d\nu = I_z
(1+z)^{-\alpha+1} \\ 
I_z & = & (1+z)^l \int_{\nu_1}^{\nu_2} C_0 \nu^{-\alpha} d\nu \\ 
I_z & = & I_0 (1 + z)^l  \label{eq:levol}
\end{eqnarray}
Thus, $I(z)$ is the peak luminosity (photons sec$^{-1}$) measured at
redshift z which will be in the detector bandwidth at z=0. $I_z$
includes the term $(1+z)^l$, which we use to make an accounting of
evolution in the peak luminosities of GRBs as a function of redshift,
and $I_0$ is a constant.  This form of evolution is here adopted as a
convenience, and is not motivated either by observations of evolving
populations (\eg\  radio galaxies, faint blue galaxies) nor by
theoretical predictions.  This limits the usefulness of the
conclusions to describe only those models which can be accommodated by
power-law evolution over the range of redshifts the model implies GRBs
are observed by BATSE.

To find the measured luminosity, we must correct for time dilation, by
dividing I(z) by an additional redshift factor (1+z).  The number of
photons measured at redshift z=0 will be equal to the photon
luminosity at z=0, multiplied by the ratio (f) of the solid angle
subtended by the detector (of area A) to the total solid angle (4$\pi$
ster) (Weinberg 1972, p. 421):
\begin{equation}
\label{eq:frac}
f = \frac {A}{4 \pi R_0^2 r_z^2 } 
\end{equation}
This measured flux (in phot \percm~\persec) is:
\begin{equation}
\label{eq:flux}
F(z)= \frac{I_0}{4\pi} \frac{(1+z)^{-\alpha +l}}{ R_0^2r_z^2}
\end{equation}
\noindent where $I_z$ is in photons sec$^{-1}$ and  $r_z$, the
co-ordinate distance, is (from Weinberg 1972, p. 485): 
\begin{equation}
\label{eq:r}
r_z = \frac{zq_0 + (q_0 - 1)(\sqrt{2q_0z + 1} -1 )}{H_0R_0q_0^2(1+z)}
\end{equation}
where $q_0$ is the deceleration parameter, $H_0$ is the Hubble
parameter at the current epoch, $R_0$ is the expansion parameter at the
current epoch.    

We find the total number of GRBs observable out to a redshift $z'$
during an observation period of $\Delta t$ at the present epoch:
\begin{equation}
\label{eq:int}
N_b(z')= \int_0^{z'} \frac{\Delta t  }{1+z} 4\pi n_z r_z^2
\frac{dr_z}{dz} \frac{dz}{\sqrt{1- kr^2}}
\end{equation}
where $k=-1$, 0, or 1 for an open, critical, and closed universe, and
$n_z$ is the number of GRBs per comoving volume per comoving time
(frequency density) as a function of redshift z, accounting for
evolution, which we take simply to be of a law form:
\begin{equation}
\label{eq:nz}
n_z = n_0 ( 1 + z)^p
\end{equation}
\noindent with $n_0$ a constant.   When p=0, the model is one of no
evolution; when p$>$0, then we have ``positive'' evolution -- an
increasing comoving burst rate with increasing redshift -- and when
p$<$0 we have ``negative'' evolution -- a decreasing comoving burst rate
with increasing redshift.  Eq.~\ref{eq:int} can be rewritten as:
\begin{equation}
\label{eq:intnok}
N_b(z') = \Delta t  \frac{4 \pi n_0}{q_0^4 (R_0 H_0)^3} \int_0^{z'} \frac{ (1+z)^{p-4} \left(z q_0 +
(q_0-1)[\sqrt{1 + 2 q_0 z} - 1] \right)^2 }{\sqrt{1 + 2 q_0
z}}~dz
\end{equation}

Using these assumptions, we perform the following analysis.  We define
a sample of M GRB peak fluxes, taken from the second BATSE catalog
(\cite{fishman94}), from bursts which have a peak flux measured on the
1024 msec time scale greater than 0.3 photons \percm \persec in the
100-300 keV band, the peak flux to which the trigger efficiency is
confidently known (C. Meegan, private communication).  We find 394
bursts which meet these criteria.

We sort the peak fluxes in increasing order and produce an
observational integrated number -- peak flux curve.  We correct the
observational number -- peak flux curve for the BATSE trigger
efficiency table, given in the second catalog.  We use an analytic
approximation to the trigger efficiency of the form:
\begin{eqnarray}
{\rm efficiency}=A - B \exp^{\frac{C-F_{\rm peak}}{D}} \label{eq:eff} \\
A &= &0.9989 \nonumber \\
B &= &0.003933 \nonumber \\
C &= &0.6496 \nonumber \\
D &= &0.0748 \nonumber 
\end{eqnarray}
and $F_{\rm peak}$ is the peak flux measured by BATSE.  This equation
is accurate to better than 1.2\%, which is sufficient accuracy for
the burst samples sizes considered here.

We assume a value for \oo(=2$q_0$), which sets the functional
dependence of $r_z$ on z (eqn.~\ref{eq:r}).  We then:
\begin{enumerate}
\item assume a value for \zmax, the redshift from which the burst with
lowest observed peak flux was emitted ($F(z_{\rm max})$), and numerically
integrate eqn.~\ref{eq:int} to find $N_b(z_{\rm max})$;
\item set $N_b(z_{\rm max})$=M (here, 394).  This sets the value $n_0$ as a function of
assumed \zmax~ and $q_0$, with knowledge of $H_0$. 
\end{enumerate}
Then, for each burst, ordered in increasing peak
flux, we do the following:
\begin{enumerate}
\item from eqn.~\ref{eq:flux}, we find the ratio of burst peak flux to
the minimum peak flux of a detected GRB to be:
\begin{equation}
\label{eq:ratio}
\frac{F(z)}{F(z_{\rm max})} = \left(\frac{r_{z_{\rm max}}}{r_{z}}\right)^2
\left(\frac{1+z}{1+z_{\rm max}}\right)^{-\alpha + l}
\end{equation}
This sets the value $z$ at which the burst flux $F(z)$ was measured.
The ratio of $r_z$ with $r_{z_{\rm max}}$ is independent of $n_0$,
$H_0$ and $R_0$ -- thus, dependent only on z, $\alpha$, and l. 
\item Using $z$ in eqn.~\ref{eq:int}, we find $N_b(z'(F))$, which
permits us to produce a theoretical \lnls~ distribution, for an
assumed \zmax, $\alpha$, and \oo, l, and p. 
\end{enumerate}

After producing a theoretical \lnls~distribution as we describe, we
compare this to the observed distribution, using a single distribution
KS test, to find the probability that the observed distribution was
produced by the model distribution.   

This was done for 3 values of $\alpha$, and for $0.1\le$\zmax$\le 955$
at points separated by $\Delta$~log(\zmax)=0.041, and for $10^{-2}\le$
\oo~$\le 2$, separated by  $\Delta$~log(\oo)=0.041.

The calculation was performed for the case of no frequency density or
peak luminosity evolution (p=0, l=0), and separately for frequency
density evolution (integer values of p between -4 and 4, inclusive)
and peak luminosity evolution (integer values of l between -4 and 4,
inclusive).

\subsubsection{Procedure for Complex Spectrum}

The procedure to fit the \lnls~curve for the assumed complex spectrum
is similar to that followed above, differing only in detail.  The
single power-law spectrum is responsible for the simple analyticity of
eqn.~\ref{eq:ratio}, the ratio of the peak fluxes of standard candle
bursts at different redshifts.  The complex spectrum requires that the
ratio be found numerically.  A further systematic complication is
that, as the origin of the spectral ``break'' is not well understood,
exactly at what co-moving energy the spectral break should be placed
is not obvious, although the observational limit that the vast
majority of the spectral breaks have \ebreak\ $<400$ keV
(\cite{band93}) provides an important guide.  We have made the choice
to set the co-moving breaking energy for each simulation so that the
observed breaking energy is at the top of the passband of the 2B
catalog peak fluxes (300 keV) for a burst at the {\it implied} lowest
redshift \zmin\ (that is, for the brightest burst).  Thus, for a model
with \zmax\ which requires a certain \zmin, then the assumed spectrum has
\ebreak=300(1+\zmin)~keV (found iteratively) and therefore for all
fainter bursts the \ebreak\ (observed at earth) will lie between 300 --
300(1+\zmin)/(1+\zmax) keV, as breaking energy is held constant.  We
discuss the implications of this choice further in
Sec.~\ref{sec:discussion}.

As in Eq.~\ref{eq:flux}, the measured flux for the complex spectrum
burst is:
\begin{equation}
\label{eq:fluxcomp}
F(z)= \frac{ \int_{\nu_1(1+z)}^{\nu_2(1+z)} \frac{dI_\nu}{d\nu} d\nu}{ 4\pi R_0^2r_z^2}
\end{equation}
\noindent where the differential photon spectrum is given in
Eqs.~\ref{eq:bandspectra1}-\ref{eq:bandspectra2}, and the values $\alpha_b$=1.0 and
$\beta$=2.5 are adopted.  Thus, the ratio of burst peak flux to the
minimum peak flux of a detected GRB is (replacing eq.~\ref{eq:ratio}):
\begin{equation}
\label{eq:complexratio}
\frac{F(z)}{F(z_{\rm max})} = \left( \frac{r_{z_{\rm max}}}{r_z}
\right)^2   \left( \frac{ \int_{\nu_1(1+z)}^{\nu_2(1+z)}
\frac{dI_\nu}{d\nu} d\nu}{ \int_{\nu_1(1+z_{\rm max})}^{\nu_2(1+z_{\rm
max})} \frac{dI_\nu}{d\nu} d\nu} \right)
\end{equation}

The value of \ebreak\ (=$h\nu_b$; Eq.~\ref{eq:bandspectra1} \& \ref{eq:bandspectra2}) is found by
iteratively F(\zmin)/F(\zmax) (=ratio of the highest to lowest
observed GRB peak flux) until \ebreak/(1+\zmin)~=~300~keV.  The
integral is evaluated with the limits $h\nu_1= 50 {\rm keV}$ and
$h\nu_2 = 300$ keV (the passband in which the peak flux is measured in
the 2B catalog). 

Once the value of \ebreak\ is set, the analysis proceeds using
Eq.~\ref{eq:complexratio} to find the value $z$ at which the burst
flux $F(z)$ was measured.

\subsection{Observational confrontation with cosmological model}

An essential parameter to this model is \oo.  Using the zero-evolution
and ``standard candle'' luminosity assumptions, we estimate how many
GRBs are required to differ between parameter values of \oo~ with a KS
probability of 1\%, assuming a detector which is sensitive to GRBs out
to a redshift \zmax.  Our motivation for doing so is that this
parameter is the least constrained by the present analysis of those in
the considered model.  A calculation of how many GRBs are required to
constrain the parameter, based on these very (and probably, overly)
simple assumptions, would set an upper limit to the number of GRBs
required for the intrinsic characteristics of the source population
(\eg\ luminosity function, frequency density evolution, luminosity
evolution) to manifest themselves.

We wish to estimate the number of GRBs (N) required to discern between
two values of \oo, assuming standard candle GRBs of constant number
per comoving time per comoving volume.  To do this, we assume a
\zmax, the redshift at which the standard candle GRBs produce a flux
at the completion limit.

We produce \lnls~distributions for \oo=1.0 and
\oo=0.1, assuming a constant number of bursts per comoving time
per comoving volume, sampled at 200 points separated by a constant
amount of comoving time-volume.  The normalized distributions are the
cumulative probability distributions, and are not a function of the
number of detected bursts.  We compare the two distributions using a
single distribution KS test. 

For those cases when the number of required GRBs was less than the
sampling of the cumulative probability distribution, the process was
repeated, with the cumulative probability distributions sampled,
instead of with 200 GRBs, with N GRBs found in the first iteration.

We perform the same analysis to estimate the number of GRBs required
to discern between values of \oo=0.5 and \oo=0.1.
                                                       
\section{Results}

\subsection{Number -- Peak flux distribution: no evolution}

The results of fitting the peak photon count rate number distributions
to model populations with no evolution (p=0, l=0) are shown in
Fig.~\ref{fig:ndist}.  Each panel indicates the spectrum assumed
(``complex'' or single power-law $\alpha$=1.0, 2.5).  The lines are
constant KS probability (1\%, 10\% and 33\%) that the modeled
number-peak flux distribution produces a distribution as or more
disparate than the observed distribution.

The most probable values of \zmax, evaluated at (\oo=1) are
$\sim$1--2.2 (depending on the assumed spectrum). For all spectra
considered, the 90\% confidence limits on \zmax~are 0.8-3.0 (at \oo=1)
and are largely insensitive to the assumed value of \oo.  As these
confidence ranges assume the source population to be entirely of a
single spectral type, the confidence range is systematically
overestimated.

As expected, the behavior of the complex spectral model as a function
of (\oo, \zmax) is qualitatively similar to that of the single
power-law spectra models, and the resulting acceptable parameter space
is between that of the $\alpha=1.0$ and 2.5 models.  We therefore
proceed with confidence that the behavior of the two single power-law
models will place conservative limits on the behavior of the complex
spectrum model.  It is desirable to do so only because of the
computational demands of the complex spectral model.

Each of the comparisons of theoretical curves with the observed curves
results in a value of the constant $n_0$.  This value is the necessary
normalization to obtain the number of observed GRBs M during the
observation period out to \zmax, from Eq~\ref{eq:intnok}.  

Following the procedure used by MP (Eq 13-15), we renormalize $n_0$ to
\nstar, the number of GRBs per \lstar~galaxy per $10^6$ years at the
current epoch.  In Fig.~\ref{fig:ndist}, the lines marked ``10'',
``1.0'', and ``0.1'' indicate \nstar\ in units of GRB per~\lstar~galaxy
per~$10^6$~yr.  For all spectra considered, the value of \nstar~ is
conservatively in the range 0.3-5.0 per \lstar~galaxy per $10^6$ yr at
\oo=1 (99\% confidence), consistent with values found by other
investigators (\cite{dermer92}; MP; \cite{wick93}).

For high \zmax, one might reasonably expect that there is evolution in
the luminosity density of galaxies.  The re-normalization (to \nstar)
we adopted here is thus pertinent only for redshifts where such
evolution can be neglected.  To change the values back to burst rate
per comoving volume at the present epoch, one multiplies by the
galaxy luminosity density we used (from \cite{efst88}):
\begin{eqnarray}
n_0 &=& n_* \times \phi^* \Gamma (2 + \alpha_g)\\
\phi^* \Gamma (2 + \alpha_g) &\approx&1.67 \times 10^{-2} L^*
h_{100}^3 {\rm Mpc}^{-3}
\end{eqnarray}
\label{sec:nstar}
As this re-normalization is multiplication by a constant, the {\it
relative} burst frequency densities do not change; thus \nstar~may be
used to compare frequency densities beyond redshifts where evolution
becomes important, with the assumption that at such redshifts the
burst frequency densities are unrelated to galaxy evolution.

\subsection{Observational confrontation with cosmological model}

As is evident in Fig.~\ref{fig:ndist}, the Number distributions are
largely insensitive to the value of \oo, for the 394 GRBs in this
sample.

In Fig.~\ref{fig:qcomp}a \& b, we show the number of GRBs which would
be required to discern between \oo=1 and 0.1, \& \oo=0.5 and 0.1, from
the number-peak flux distribution, using data which is complete for
standard candle GRBs out to a redshift of \zmax.

The points are the results of the calculation at a few values of
\zmax, assuming different spectra ($\alpha$=1.0 or 2.5); the lines
between the points are there to ``guide the eye'', and also identify
the spectral slope assumed for the calculation.  The solid line is
$\alpha$=1. and the dashed line is $\alpha$=2.5.

The most likely \zmax~for BATSE GRBs calculated in the previous
section was (2.2, 1.0) for $\alpha$ = (1.0, 2.5).  For both spectra,
this corresponds to $\sim$ 9,000 required for BATSE to discern between
\oo=1 and 0.1.  Similarly, approximately 20,000 GRBs are required to
discern between \oo=0.5 and 0.1.

The uncertainties on these values are large.  Permissively, if the
actual \zmax~ value is 3.0 (90\% upper limit for $\alpha$=1.0), then
only \about 4,000 (or 10,000) GRBs are required to discern between
\oo=1 and 0.1 (or 0.5 and 0.1).

The number of GRBs required diminishes by 4 orders of magnitude when
\zmax~ changes from 0.5 to 10.  For \oo=1.0, a change in depth of
\zmax=1.0 to 10.0, requires an increase in detector sensitivity of a
factor of (30, 400) for $\alpha$=(1.0,2.5).

In Fig.~\ref{fig:qcomp}c \& d, we show the coverage (in years $\times
4\pi$ ster) required for a detector complete to burst peak fluxes out
to \zmax~to obtain the number of GRBs required to discern between
different values of \oo~in the \lnls~ distribution, assuming a
constant number of GRBs per comoving time, per comoving volume. We
normalize to 394 GRBs detected with peak flux $>$ 0.3 phot
\percm~\persec on the 1024 msec integration timescale in the $2.7
\times 10^8$ sec ster coverage of the second BATSE catalog, which
introduces \about 5\% uncertainty in the coverage.

At the minimum, BATSE (assuming 2.6 $\pi$ ster coverage) would require
$\approxgt$ 6 years integration time before the parameter \oo~ becomes
reasonably constrained by the Number-- peak flux distribution to
permit confrontation the cosmological models using only the
distribution.  The most likely required integration time is \about 30
years.  Detectors sensitive out to \zmax=10 would give statistically
different Number-peak flux distributions after a few weeks of full sky
coverage, measuring $\approxlt$200 GRBs.

\subsection{Number -- Peak flux distribution: with evolution}

\subsubsection{Frequency Density Evolution}

In Figs.~\ref{fig:p1.0} - \ref{fig:p2.5}, we show the results of
comparison of the observed BATSE number -- peak flux distribution with
distributions calculated including evolution in the frequency density
of GRBs for $-4 \le p \le 4$.  As in Fig.~\ref{fig:ndist}, the solid
lines are constant KS probability (1\%, 10\% and 33\%) that the
assumed model could produce a distribution as or more disparate than
that observed. The broken lines are constant \nstar~ (see
Sec.~\ref{sec:nstar}).  In panels a, b, and h of each figure, the
constant \nstar~ lines are marked in units of GRBs per \lstar~galaxy
per $10^6$ yr and are identical to the lines in the other panels.

In the case of ``negative'' density evolution (p$<$0; panels b, d, f,
and h), stronger evolution has the effect of decreasing the implied
\zmax~ values.  If the number density decreases with increasing
\plusz, lower values of \zmax~ are required; as a result, higher
values of \nstar~ are required, increasing by a factor of $\sim$10 as
p decreases from $-1$ to $-4$.  This behavior is qualitatively the
same independent of the assumed spectrum (of those considered here).
The observed number -- peak flux distribution is consistent with
strong negative density evolution up to p=$-4$, but requires a lower
\zmax~ and a higher \nstar~ (by $\times10$) than a zero-evolution
scenario.

In the case of ``positive'' density evolution (p$>0$; panels a, c, e,
and g), stronger evolution has the effect of increasing the implied
\zmax~ values.  For p$>$2, the models could not always be reasonably
fit to the observed distribution for values of \zmax\ $<$1000, and
exhibit dependency on the assumed spectrum.  The \zmax~ values are
strongly dependent on the assumed source spectrum, with the range of
acceptable \zmax~ values increasing greatly for lower values of \oo~
($<$1).

\subsubsection{Peak Luminosity Evolution}

Figs.~\ref{fig:l1.0} - \ref{fig:l2.5} show the results of comparison
of the observed BATSE number -- peak flux distribution with
distributions calculated including evolution in the peak luminosity of
GRBs with (1+z) according to Eq.~\ref{eq:specev}.  As in
Fig.~\ref{fig:ndist}, the broken lines are constant KS probability
(1\%, 10\% and 33\%) that the observed number -- peak flux
distribution could have been drawn from the model distribution.  The
solid lines are constant \nstar.  In panel a of each figure, the
constant \nstar~ lines are marked in units of GRBs per \lstar~galaxy
per $10^6$ yr at the present epoch, and are identical to the lines in
the other panels.

In the case of ``negative'' luminosity evolution (l$<$0; panels b, d,
f, and h), stronger evolution has the effect of decreasing the implied
\zmax~ values, while increasing the \nstar~ values.  The most probable
values of \zmax~ decrease from the range of $\sim$ 1.0-2.2 (no
evolution) to $\sim$ 0.3-0.7 (p=4), and the range of implied \nstar
increases by a factor of 10 ($\sim$ 0.3-5 to 3-50 GRBs per
\lstar~galaxy per $10^6$ yr at the present epoch).

In the case of ``positive'' luminosity evolution (l$>$0; panels a, c,
e, and g), stronger evolution increases the value of \zmax, and also
constrains the value of \oo~ from above.  Thus, for models with $l >
0$, the observed number - peak flux distribution distorts the
acceptable parameter space considerably from that of zero evolution
models; the acceptable parameter space becomes a strong function of
both \zmax~ and \oo~ (see Figs~\ref{fig:l1.0}g, \ref{fig:lcmp}c, and
\ref{fig:l2.5}c).  Acceptable parameter space is found for l as high as
3, (complex spectrum or $\alpha$=2.5) (Figs.~\ref{fig:lcmp}c and
\ref{fig:l2.5}c).

\subsection{Relative Time Dilation}
For values of \zmax~and \oo~ which produced models with KS
probabilities $>$1\%, we also found the ratio \rrel: 
\begin{equation}
 R_{\rm rel}=(1+ z_{\rm max})/(1+ z_{\rm min})
\end{equation}
the ratio of the maximum redshift of GRBs in the sample to the minimum
redshift of GRBs in the modeled 394 GRB sample. This ratio is the
maximum amount of relative time dilation (or relative energy shift)
between GRBs within the observed sample.  In models with no source
evolution, this ratio is very close to (1+\zmax).  However, when
strong positive frequency density evolution is present (p$>$1), this
ratio was usually much smaller than (1+\zmax).  This is not
unexpected, as frequency density must change sharply in \plusz,
requiring that the bursts be localized within a small range of
redshifts.  For all spectra considered, the ratio \rrel\ is always
$\approxlt 9$ (for all models with \oo$<$1 and KS probability $>$10\%)
even for \zmax~ values in the range of 100-1000.

In Fig.~\ref{fig:Rrel2}, panels a, b, and c, we show the 1\% KS
probability contours for the models for each assumed spectrum with
positive frequency density evolution which still had acceptable
parameter space at \zmax $<$1000 (taken from Figs.~\ref{fig:p1.0}e,
\ref{fig:pcmp}c, and \ref{fig:p2.5}c).  In all cases, the values of
\rrel\ are smaller than (1+\zmax), indicating that all 394 observed
GRBs (in these models) come from \zmin$ > 1$.  The very shallow
spectrum $\alpha=1.0$ requires a range of \rrel\ which is relatively
high (30-60) while the more realistic complex spectrum, as well as the
$\alpha=2.5$ spectrum, has relatively low values of \rrel\ (5-8) for
models which place the faintest observed BATSE bursts at \zmax\
100-1000.

We show similar results in Fig.~\ref{fig:Rrel2}d-f for the models with
positive peak-luminosity evolution (taken from Figs.~\ref{fig:l1.0}g,
\ref{fig:lcmp}c, and \ref{fig:l2.5}c).  In all cases, the value of
\rrel\ is, at lowest, only a factor of a few lower than \zmax,
requiring large amounts of relative time-dilation for bursts at high
values of \zmax\ (factor of 20-200 for \zmax $\sim$20-800).  Models
with relatively small amounts of peak luminosity evolution (l=1 -- 2),
place the faintest GRBs at \zmax $\about 10$ with \rrel\ $\about$ 10.
 
\section{Discussion and Conclusions}
\label{sec:discussion}
We have examined the observed \lnls~distribution of 394 GRBs observed
by BATSE, and, assuming a cosmological model with a constant number of
bursts per comoving volume per comoving time, the distribution is
consistent with the ``standard candle'' GRBs with peak fluxes $>$ 0.3
phot \percm~\persec originating at \zmax=0.8-3.0 (90\% confidence), with
the most likely values in the range of 1.0-2.2, largely insensitive
to the assumed \oo.  

In performing this analysis we assumed that consistent model parameter
space of the observed GRB sample could be bracketed by the single
power-law spectral models ($\alpha$=1.0 and 2.5) which, themselves,
``bracket'' the GRB broken power-law spectra fit by Band \etal\ (1993)
(see Fig.~\ref{fig:bandspec}).  We find this assumption to be
justified, as the dominant portion of the Band \etal\ (1993) spectra
fall between these two single power-law spectral models, and the
behavior of the complex spectrum also fell between that of the two
single power-law spectral models.  We presumed that the broken
power-law parameter \ebreak\ is constant (that is, is measured to be
\ebreak/(1+z) from a burst at the redshift 1+z) and that a dominant
number of bursts have an \ebreak $\sim$ 300 keV.  In the worst
possible case of inconsistency with this assumption -- that \ebreak\ 
varies randomly -- the behavior of the models will be bracketed by
behavior of the broken power-law spectrum at its extremes, which
behaves like a single power-law spectrum of $\alpha \approxgt 1.0$ or
$\approxlt$2.5.

Our choice for the the value of \ebreak\ in our simulations was made
to maximize the effect of the difference between the ``complex''
spectrum and the power-law spectrum.   This was done to see
investigate if the power-law approximation models produce grossly
different results from the more observationally correct ``complex''
spectrum.  We find that the behavior of the ``complex'' spectrum model
is within the limits of the behavior set by the two single-power-law
spectrum models.  This is not unexpected; if the redshifts over which
the bursts span are modest in extent, then the broken power-law model
behavior is dominated by the low-end of the spectrum (\ie\
$\alpha_{\rm b}=1.0$; Eqn.~\ref{eq:bandspectra1}), while if the bursts
span over a large range of redshifts, the broken power-law model
behavior is dominated by the the high-end of the spectrum (\ie\ $\beta
= 2.5$; Eqn.~\ref{eq:bandspectra2}).

We find that, for the zero-evolution models, strong limits may be
placed on the GRB rate at at 0.3-5.0 GRBs per \lstar~galaxy per $10^6$
yr, for \oo$\approxlt$1.  This is consistent with the values found by
MP for an \oo=1 universe (although MP assumed flatter energy spectra
than has been observed by BATSE ;  $\alpha$=0.5-1.5 vs 1.5-2.5) and
with those found by Piran (1992).

When this work was largely complete, similar analyses by Cohen \&
Piran (1995) considering non-evolving cosmological models using the 2B
BATSE data base came to our attention.  Our conclusions are consistent
with their analyses. 

We find that, for BATSE, $\approxgt$4000 (most likely $\sim$ 9,000)
GRBs are required to use \lnls~to constrain a zero-evolution,
standard-candle peak luminosity cosmological model parameter \oo~ to
between 1.0 and 0.1, and $\approxgt$9000 to constrain \oo~ to between
0.5 and 0.1.  This requires a minimum of 6 yrs of BATSE integration; a
duty cycle of 50\% places this above the projected mission lifetime of
BATSE.    

The observed number -- peak flux distribution can accommodate both
peak luminosity evolution of GRBs and frequency density evolution.

The frequency density evolution, if any, can be accommodated by a
power law in $(1+z)^p$ for the ranges of \zmax~ and \oo~ considered
here.  We find no acceptable models for p$\approxgt$3.  

If we assume the difference in timescales (by $\times$2) between
bright and faint bursts found by Norris \etal~(1994a) and Wijers \&
\bog~(1994) is due to the relative time dilation of bursts within the
sample, then we find acceptable model-parameter space which would place
the faintest GRBs observed by BATSE at z as high as 1000 with maximum
relative time dilation of \rrel\ $\sim$8.  As the aforementioned studies
necessarily average over a substantial fraction of the brightest and
faintest bursts, the relative time dilation they measure should be
somewhat smaller than our parameter \rrel\ (which compares only the
brightest and faintest single bursts).  We find that values of \rrel\
as high as  $\sim$ 10  are consistent with the measurements of the
factor of two difference in timescale between bright and faint bursts.

The p=1 (frequency density evolution) model is identical to the
assumed GRB model used by \wick\ (1993) (which they proposed as a
non-evolving model).  For \oo=1 in this model, we find higher values
of the most probable \zmax~ (1.7-3.0 vs. 0.5-1.7; comparing with their
Fig. 2, at the completion limit).  This is probably due to the fact
that we use a lower flux limit (and thus, ``see'' to greater
redshifts); when we perform the same analysis to the 99\% flux limit
of BATSE (the flux limit used by \wick\ \etal), we find \zmax~ in the
range 0.9-1.7, similar to that found by \wick\ (1993).

We have found that the strongest positive frequency density evolution
scenario (p=2) with acceptable model parameters for all spectra
considered permit \zmax~ to be as great as \about 200, with values of
\rrel\ as high as 9 for hard bursts, and in the range of 2-5 for
softer bursts.  When we consider that many GRBs of different spectra
were averaged over by Norris \etal (1994a) and Wijers \& \bog~(1994)
to obtain the timescale difference of a factor of 2 between bright and
faint bursts, we estimate that the p=2 scenario cannot be excluded on
this basis; averaging together harder GRBs with softer GRBs diminishes
the higher values of \rrel, and the averaging together of many bursts
of different peak fluxes (and thus, at different redshifts) diminishes
the amount of relative timescale difference within the burst sample to
an level consistent with these observations.

Application of Eq.~\ref{eq:flux} to the p=2 scenario shows that the
implied photon luminosities in the redshifted passband at the epoch of
the source are a function of $\alpha$, \zmax, and \oo, and can be
different for different assumed spectra by several orders of magnitude
or identical to within a factor of order unity, dependent on the
assumed \oo.  For example, for an assumed spectrum of $\alpha$=1.5,
the p=2 scenario places statistically acceptable (KS prob $>$ 1\%)
populations at (\zmax, \oo) of about (10, 1.0), (60, 0.2), and (200,
0.06).  For these values of (\zmax, \oo) the ratio of photon
luminosity to flux is $4\times~ 10^{59}$, $2\times~ 10^{62}$, and
$7\times~10^{64}$ (all in units of \percm$ h_{80}^{-2}$.  The implied
source luminosity to flux ratio of non-evolving population, with a
source at \zmax\ of 1.5 for \oo=1 is $6\times~ 10^{57}$~${\rm cm}^{-2}
h_{80}^{-2}$.  Thus, the evolutionary scenarios require sources with
much greater luminosities, by up to 7 orders of magnitude) than
non-evolving scenarios.

Photons with energies as high as 10 GeV have been observed from some
GRBs (Schneid \etal 1992; Kwok \etal\ 1993; Dingus \etal 1994; Sommer
\etal\ 1994).  The Cosmic Microwave Background Radiation is capable of
scattering photons at such energies when they are emitted from very
high redshifts (\cite{babul87}).  There is essentially a ``wall'' in
redshift-space to photons of a given energy due to scattering by the
CMBR, such that photons of energy $E$ MeV will not be observed if they
are emitted beyond a redshift $z \approx 7400 (E/{\rm 1
  MeV})^{-0.484}$ (\cite{babul87}).  Thus, photons of energy 1 GeV
would not be observed from a GRB if the GRB occurred at a redshift
greater than \about 260.  Photons of such energies have been detected
from only a very small fraction of the observed GRBs, and these GRBs
are typically among the brightest observed, which may come from
redshifts a factor of \about 5-8 lower than the maximum redshift; for
example, if \zmax=1000, the brightest GRBs would come from redshifts
125-200, and would thus these high energy photons would not be
scattered by the CMBR.  Thus, the observation of some such photons is
not inconsistent with BATSE seeing GRBs from redshifts as great as
1000.    

The implied comoving burst rate in the strong (p=2) frequency density
evolution scenario is $\times 100$ higher for hard bursts
($\alpha$=1.5) than for soft bursts ($\alpha$=2.5).  For weaker (p=1)
frequency density evolution, the comoving burst rate of harder GRBs is
only \about $\times$10 greater than that of the softer GRBs.  These
high spectrally-dependent differences in the relative comoving burst
rates are a constraint on models which contain strong GRB frequency
density evolution.  Further, `standard-candle' models which contain
strong frequency density evolution require that the GRBs of different
spectra either have different number--peak flux relations, or that the
relative burst rates conspire with the evolution rate and space-time
geometry in the epoch where the majority of GRBs of a particular
spectrum occur to produce identical number--peak flux distributions,
independent of the intrinsic source spectrum.  In
Fig.~\ref{fig:reldist}, we show the cumulative distributions of the
100 spectrally hardest and 100 spectrally softest GRBs from the
present sample; a 2-distribution KS test shows them to be
statistically identical (KS probability 59\%).

The peak luminosity evolution, if any, can be accommodated by a power
law in $(1+z)^l$.  For the peak luminosity evolution models, the
relative time dilation between the faintest and brightest bursts
(\rrel) can be as high as 200 for the models considered here, and is
very roughly between (1+\zmax)/4 and (1+\zmax), favoring the high
value for low \zmax, but decreasing to the low value for high
\zmax.  With the estimate that \rrel\ must be $\approxlt 10$ to
be consistent with the observed amount of relative time dilation
within the GRB sample, then the present analyses excludes luminosity
evolution models  which place the faintest observed GRB in the present
sample at \zmax~ $\approxgt$ 10.  If we also require, as we have in the
frequency density evolution, that the luminosity evolution scenarios
permit relative time dilation of at least a factor of 2, then we find
that that negative luminosity evolution cannot be stronger than l$=2$.

We conclude that the Number-peak flux distribution of GRBs observed by
BATSE is consistent with a homogeneous, zero evolution source
population, to a \zmax~ most probably in the range 1.0-2.0 for GRBs
with peak luminosities of \about 0.3 phot \percm~\persec, and that the
flattening observed in this distribution is consistent with being due
only to cosmological effects.  The amount of frequency density
evolution of the form given by Eqn.~\ref{eq:nz} is constrained, with
the exponent p $<$2.  The amount of peak luminosity evolution of the
form given by Eqn.~\ref{eq:levol} is constrained, with the exponent l
$<$ 2.  

As the constraints on \zmax, p, and l, improve with greater numbers of
GRBs, we expect similar future work with larger samples to be useful
in the parameterization of the GRB source population.  However, to put
practical limits on GRB cosmological models, much deeper observations
are required.  

We wish to draw the reader's attention to the fact that in considering
one evolution model (frequency density evolution or luminosity
evolution), that the other evolutionary model was held to be zero,
which may not be the actual observational case.  Given the wide range
of acceptable parameters for even this case, it is not at all clear
that it is, even in principle, possible to jointly constrain all these
different functions and values (frequency density evolution,
luminosity density evolution, \zmax, \oo, \nstar) with the single
observational \lnls\ curve.  We also wish to point out that, as we
have assumed power-law forms of the evolution, the validity of these
conclusions is limited to actual evolution which follows power-law
form.

If flattening of this distribution from the expected (from Euclidean
geometry and spatial homogeneity) -3/2 power law are due only to
cosmological effects, then BATSE will not integrate sufficient numbers
of GRBs with peak flux 0.3 phot \percm \persec to constrain \oo~ to
the range 0.1--1.0 during its projected lifetime, which would, with
certainty, permit confrontation with cosmological models.  However,
Hakkila \etal (1994) estimates that BATSE can integrate enough GRBs in
its lifetime to permit confrontation with a variety of galactic models.
A mission with much fainter flux limits ($\times$ 70-400),
fortuitously pointed toward M31, could permit both after integrating
for a period of months.

We find that the results of the present analysis is supporting, but
not compelling,  evidence of a cosmological origin for GRBs and much
further work is required to investigate this hypothesis.

RR gratefully acknowledges useful discussions with D. Buote, helpful
comments on an early version of this paper by S. Mao and T.  \wick,
and useful comments by C. Meegan and C. Kouveliotou.  We are grateful
to B. \bog\ for useful comments and suggestions.  This work was
supported under NASA Grant \#NAGW-3234 and \#NGT-513-69.  This
research has made use of data obtained through the Compton Gamma Ray
Observatory Science Support Center Online Service, provided by the
NASA-Goddard Space Flight Center.

\newpage

\begin{figure}
\caption{ \label{fig:alphahist}
Number vs.  photon number power law slope $\alpha$; for 30 GRBs.  Data
was taken from Schaefer \etal (1994); bursts were selected for this
sample which had peak photon fluxes greater than 4 photons
\percm~\persec~on the 64msec timescale -- however, some
``interesting'' bursts were also included.  }
\end{figure}

\begin{figure}
\caption{ \label{fig:bandspec}
(a) The spectral models fit by Band \etal (1993) to 55 GRBs,
normalized at 50 keV.  We show these models for the energy range
50-5000 keV, although the BATSE GRB spectra largely span 20-2000 Mev,
to show the extrapolated behavior of the spectra at higher
energies. (b)  Single power-law spectra with $\alpha$=1.0, and 2.5,
and the complex spectrum (see text), normalized at 50 keV.  These
spectra define the limiting and average behavior of the observed
spectra found by Band \etal (1993) }
\end{figure}

\begin{figure}
\caption{ \label{fig:ndist} Results of comparisons of the BATSE
observed number -- peak flux curves with that expected from a constant
number per comoving time per comoving volume source distribution,
assuming different source photon spectra (complex, or single photon
power law $\alpha$=1.0, 2.5; marked in each panel), as a function of
\oo, and the redshift of the faintest bursts observed (\zmax).  The
solid lines are constant KS probability (1\%, 10\%, and 33\%; note
that lower KS probability lines enclose higher KS probability lines --
the outer two lines are 1\% and the inner two lines are 33\%).  The
broken lines are the constant \nstar, the number of bursts per
\lstar~galaxy per $10^6$ years at the current epoch, and are marked
with their value.}
\end{figure}

\begin{figure}
\caption{ \label{fig:qcomp} Estimations of the required number of GRBs
to differentiate (at the 99\% confidence level) between two different
values of \oo, assuming standard candle GRBs and a constant number of
GRBs per comoving time per comoving volume, as a function of the value
\zmax, the redshift at which the standard candle GRB would produce a
flux equal to the completion limit, assuming no evolution in the GRB
population.  Solid line is $\alpha$=1.0, broken line is $\alpha$=2.5.
a) Number of GRBs needed to differentiate between \oo=1.0 and 0.1, as
a function of the \zmax. b) Number of GRBs needed to differentiate
between $\Omega_0=0.5$ and 0.1 as a function of \zmax~c) Coverage (in
years 4 $\pi$ ster) required to integrate the number of GRBs needed to
differentiate between \oo=1.0 and 0.1. d) Coverage (in years 4 $\pi$
ster) required to integrate the number of GRBs needed to differentiate
between \oo=0.5 and 0.1.  }
\end{figure}

\begin{figure}
\caption{ \label{fig:p1.0}
Results of comparisons of the BATSE observed number -- peak flux
curves with that expected from a parent population evolving in number
per comoving volume according to Eq.~{\protect \ref{eq:nz}}, assuming
a source power law photon index slope of 1.0, as a function of \oo,
and the redshift of the faintest bursts observed (\zmax).  The solid
lines are constant KS probability (1\%, 10\%, and 33\%; note that
lower KS probability lines enclose higher KS probability lines -- the
outer two lines are 1\% and the inner two lines are 33\%).  The
``wiggly''ness of the lines is due to the finite resolution of the
probability calculation, and does not reflect real structure in the
probability contours.  The broken lines are constant in \nstar, the
number of bursts per \lstar~galaxy per $10^6$ years at the current
epoch, and are marked with their value. }
\end{figure}

\begin{figure}
\caption{ \label{fig:pcmp} This is identical to Fig.{\protect
\ref{fig:p1.0}}, except a ``complex'' photon spectrum is assumed (see
text).}
\end{figure}

\begin{figure}
\caption{ \label{fig:p2.5}
This is identical to Fig.{\protect \ref{fig:p1.0}}, except a photon
power law spectral index of 2.5 is assumed.}
\end{figure}

\begin{figure}
\caption{ \label{fig:l1.0}
Results of comparisons of the BATSE observed number -- peak flux
curves with that expected from a parent population evolving in peak
luminosity according to  Eq.~{\protect \ref{eq:peaklum}}, assuming a source
power law photon index slope of 1.0, as a function of \oo, and the
redshift of the faintest bursts observed (\zmax).  The solid lines
are constant KS probability (1\%, 10\%, and 33\%) that the observed
Number -- peak flux distribution can be drawn from the simulated
distribution.  The broken lines are constant in \nstar.  }
\end{figure}

\begin{figure}
\caption{ \label{fig:lcmp} This is identical to Fig.{\protect
\ref{fig:l1.0}}, except a ``complex'' photon spectrum is assumed (see
text).}
\end{figure}

\begin{figure}
\caption{ \label{fig:l2.5}
This is identical to Fig.{\protect \ref{fig:l1.0}}, except a photon
power law spectral index of 2.5 is assumed.}
\end{figure}

\begin{figure}
\caption{ \label{fig:Rrel2} This figure shows the range of acceptable
values of \rrel\ for selected models with KS probabilities $>$1\%. The
broken lines are constant KS probability enclosing models at $>$1\%.
The two solid lines in each panel are marked constant \rrel. The value
of \rrel changes monotonically between the two solid contours.  Each
panel indicates the assumed spectrum and the model parameter which is
different from zero($p$ for frequency density evolution, $l$ for
peak luminosity evolution).  }
\end{figure}

\begin{figure}
\caption{ \label{fig:reldist}
Comparison of the cumulative distribution for the 100 hardest
(determined by the ratio
of the 100-300 keV photon fluence to the 50-100 keV photon fluence)
and 100 softest GRBs with 1024msec peak photon flux $>$ 0.3 phot
\percm~\persec.  They are statistically identical, with a KS
probability of 0.59 }
\end{figure}

\clearpage
\pagestyle{empty}
\begin{figure}
\PSbox{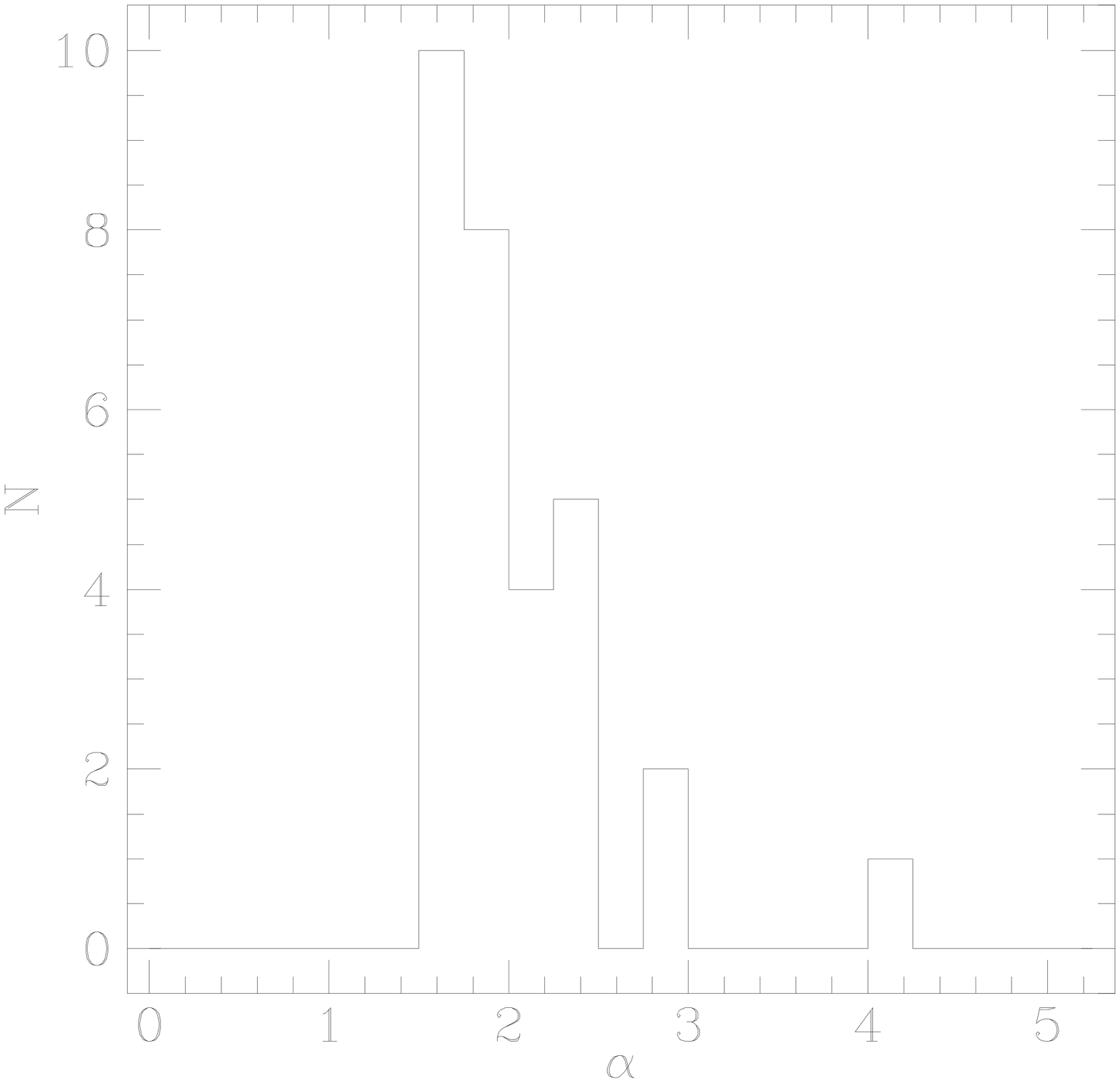 hoffset=-60 voffset=-56}{14.7cm}{21.5cm}
\FigNum{\ref{fig:alphahist}}
\end{figure}

\clearpage
\pagestyle{empty}
\begin{figure}
\PSbox{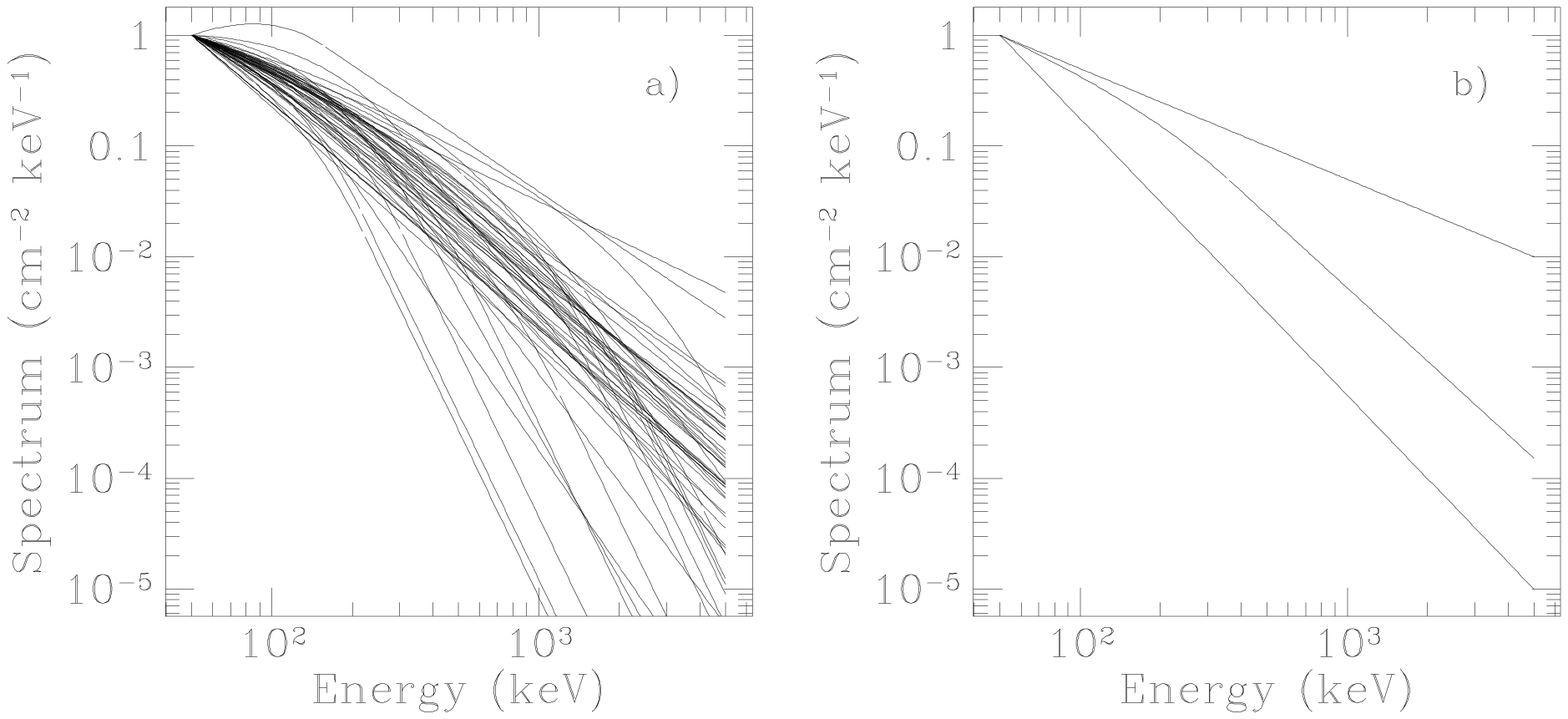 hoffset=-60 voffset=-56}{14.7cm}{21.5cm}
\FigNum{\ref{fig:bandspec}}
\end{figure}

\clearpage
\pagestyle{empty}
\begin{figure}
\PSbox{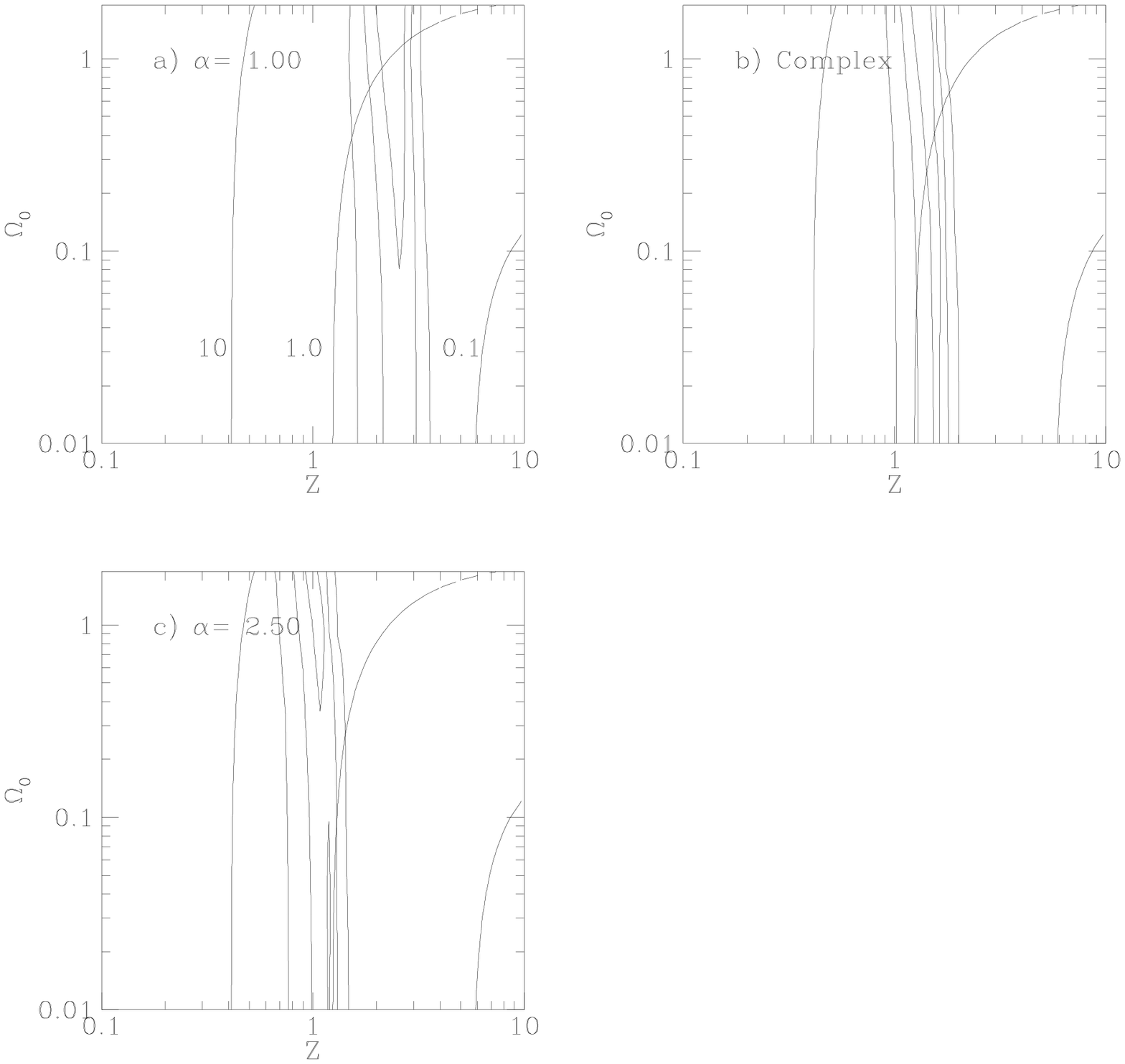 hoffset=-60 voffset=-56}{14.7cm}{21.5cm}
\FigNum{\ref{fig:ndist}}
\end{figure}

\clearpage
\pagestyle{empty}
\begin{figure}
\PSbox{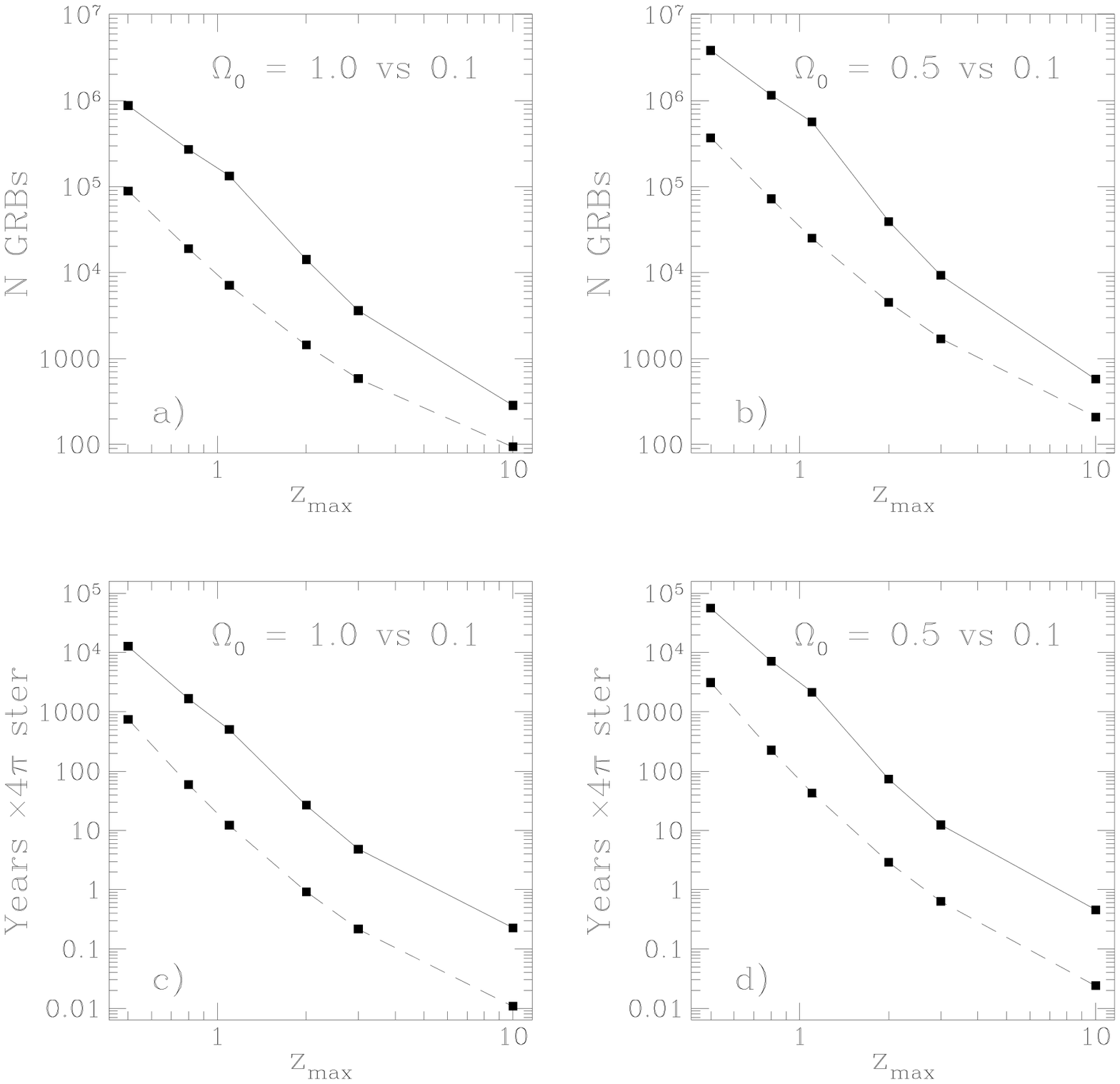 hoffset=-60 voffset=-56}{14.7cm}{21.5cm}
\FigNum{\ref{fig:qcomp}}
\end{figure}

\clearpage
\pagestyle{empty}
\begin{figure}
\PSbox{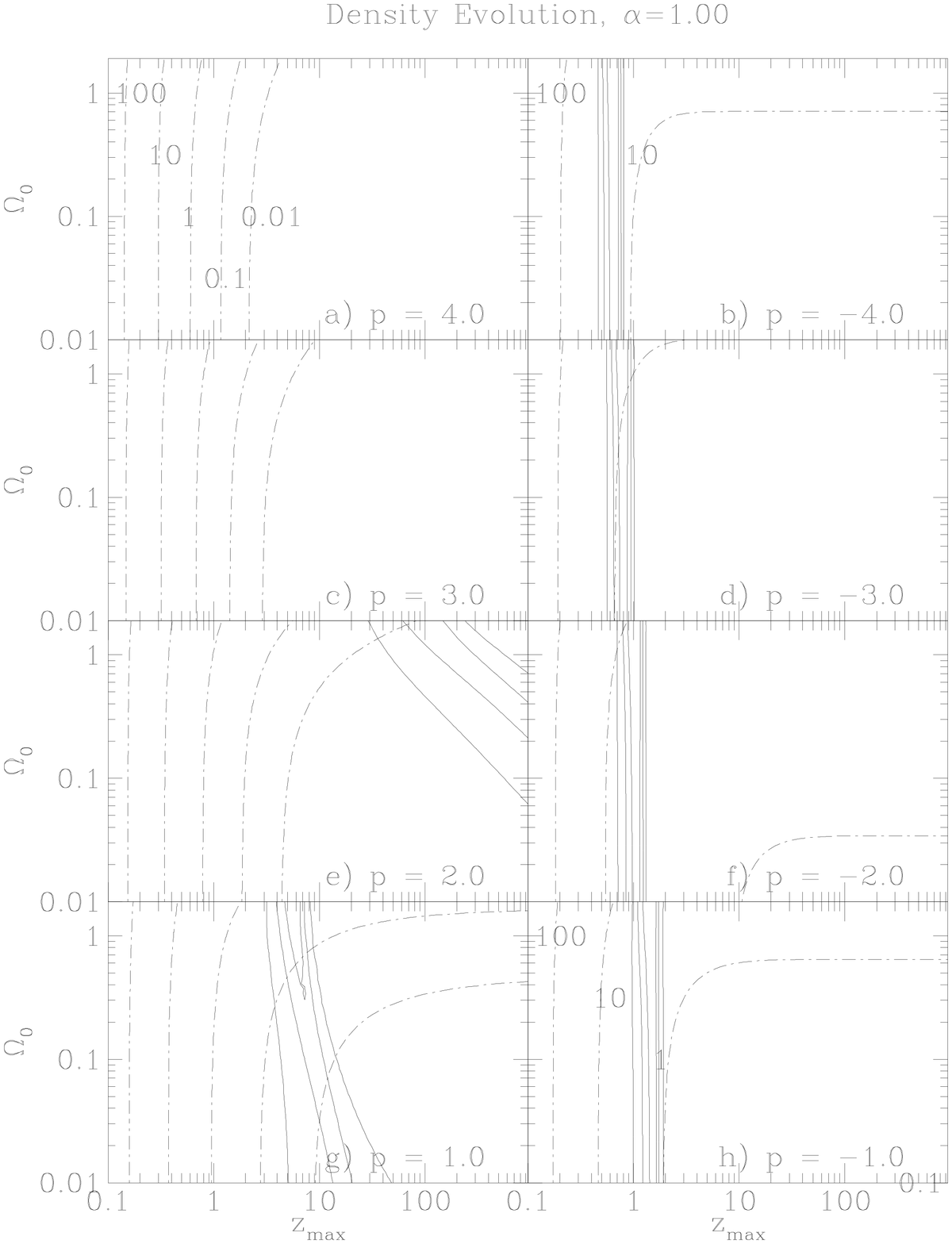 hoffset=-60 voffset=-66}{14.7cm}{21.5cm}
\FigNum{\ref{fig:p1.0}}
\end{figure}

\clearpage
\pagestyle{empty}
\begin{figure}
\PSbox{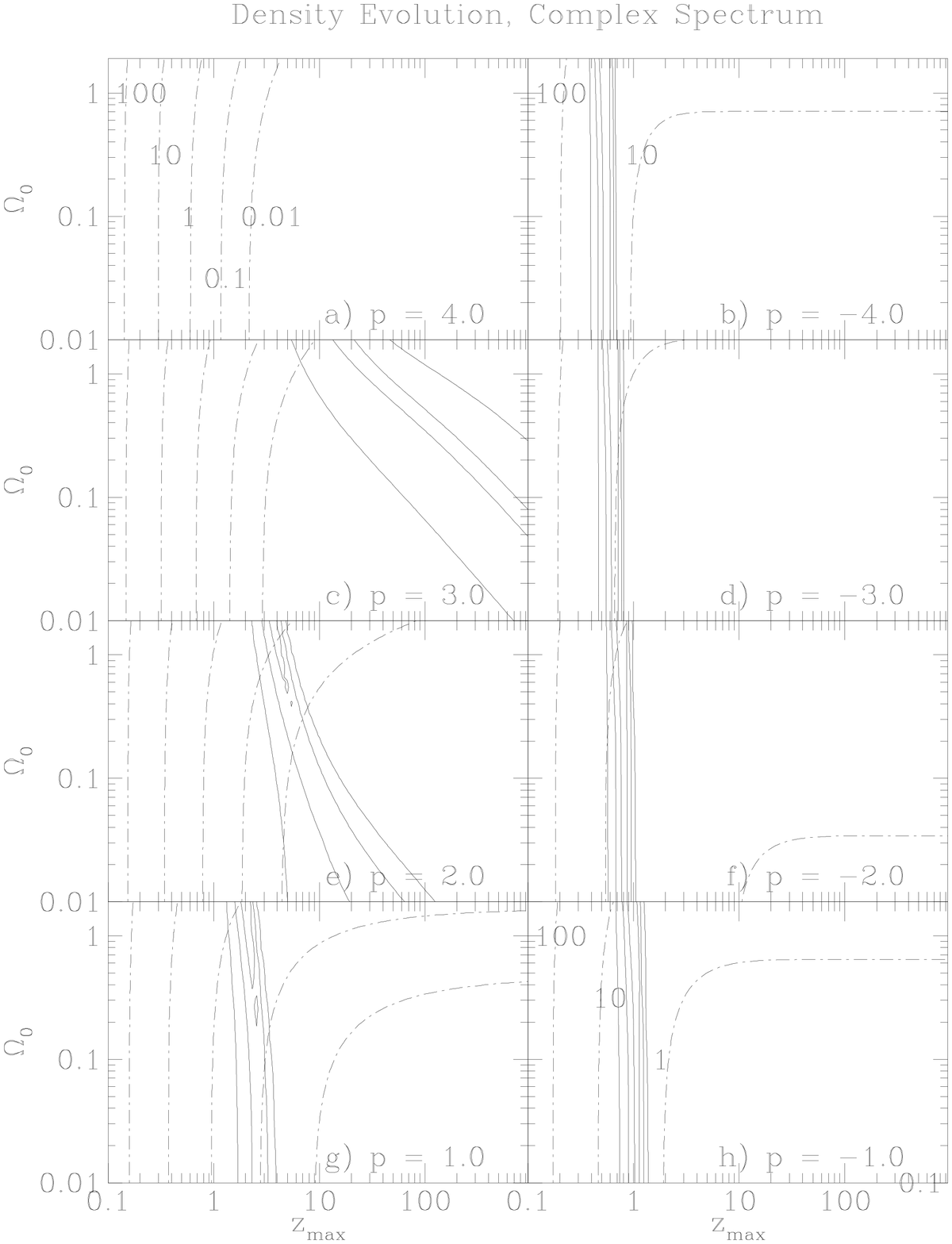 hoffset=-60 voffset=-66}{14.7cm}{21.5cm}
\FigNum{\ref{fig:pcmp}}
\end{figure}

\clearpage
\pagestyle{empty}
\begin{figure}
\PSbox{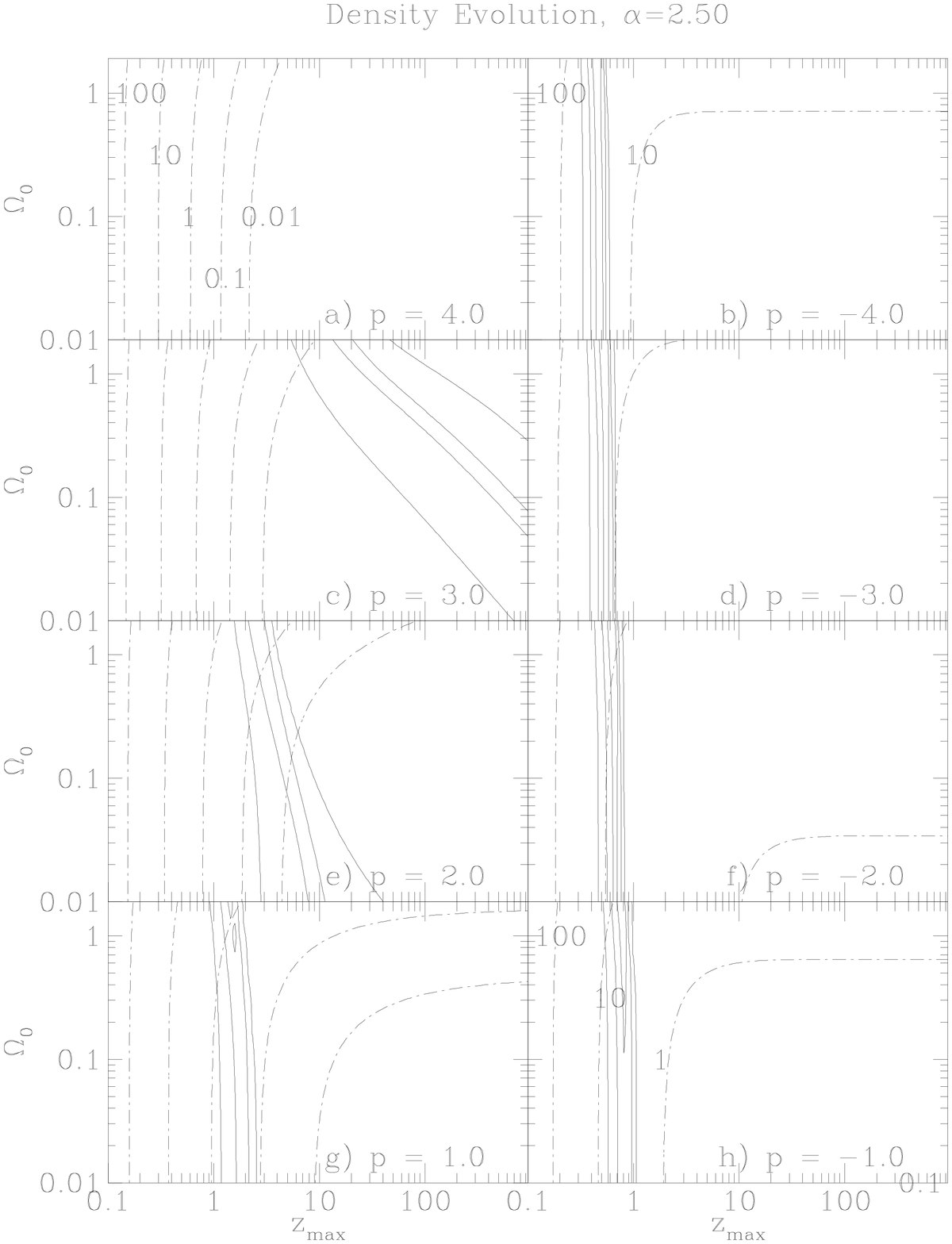 hoffset=-60 voffset=-66}{14.7cm}{21.5cm}
\FigNum{\ref{fig:p2.5}}
\end{figure}

\clearpage
\pagestyle{empty}
\begin{figure}
\PSbox{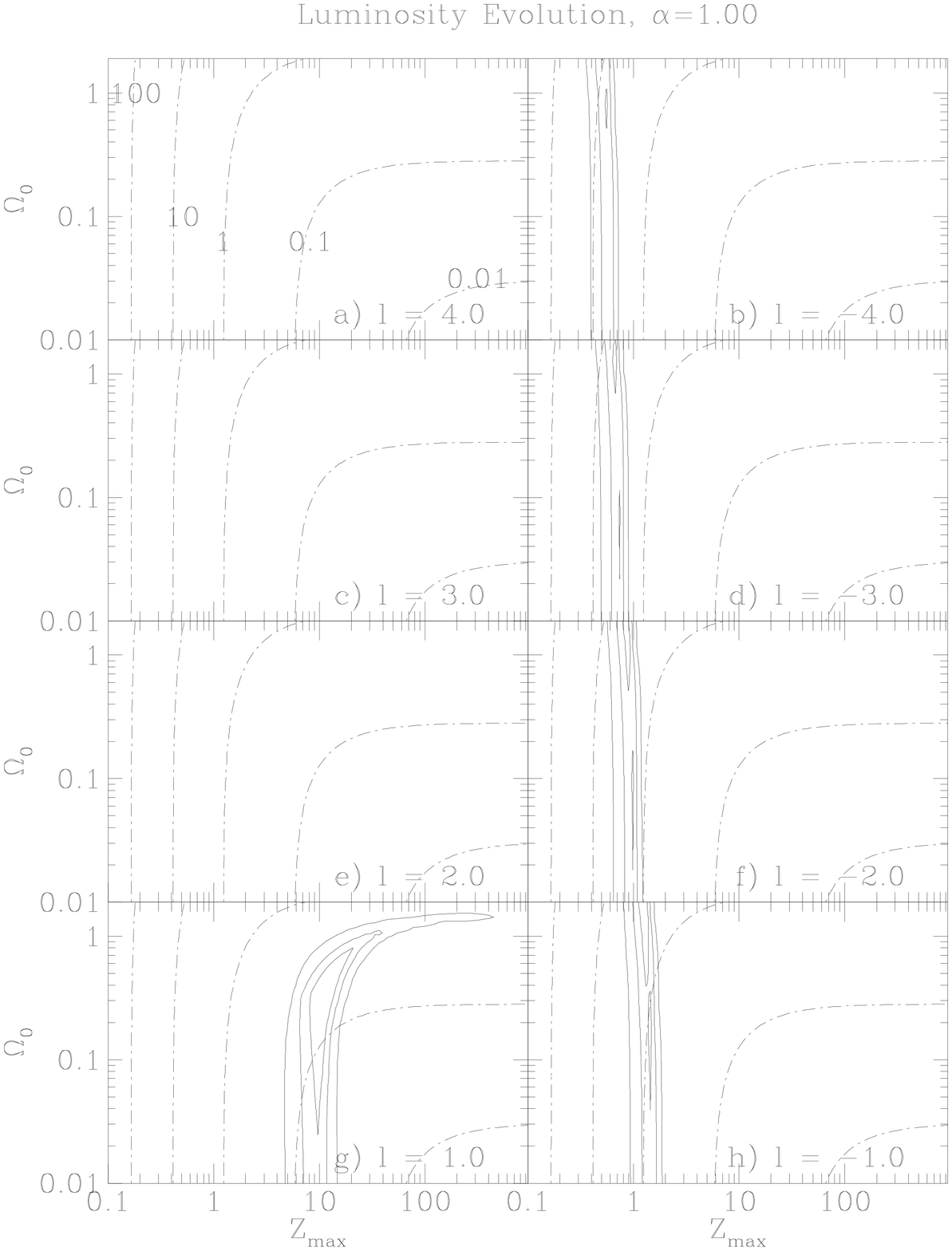 hoffset=-60 voffset=-66}{14.7cm}{21.5cm}
\FigNum{\ref{fig:l1.0}}
\end{figure}

\clearpage
\pagestyle{empty}
\begin{figure}
\PSbox{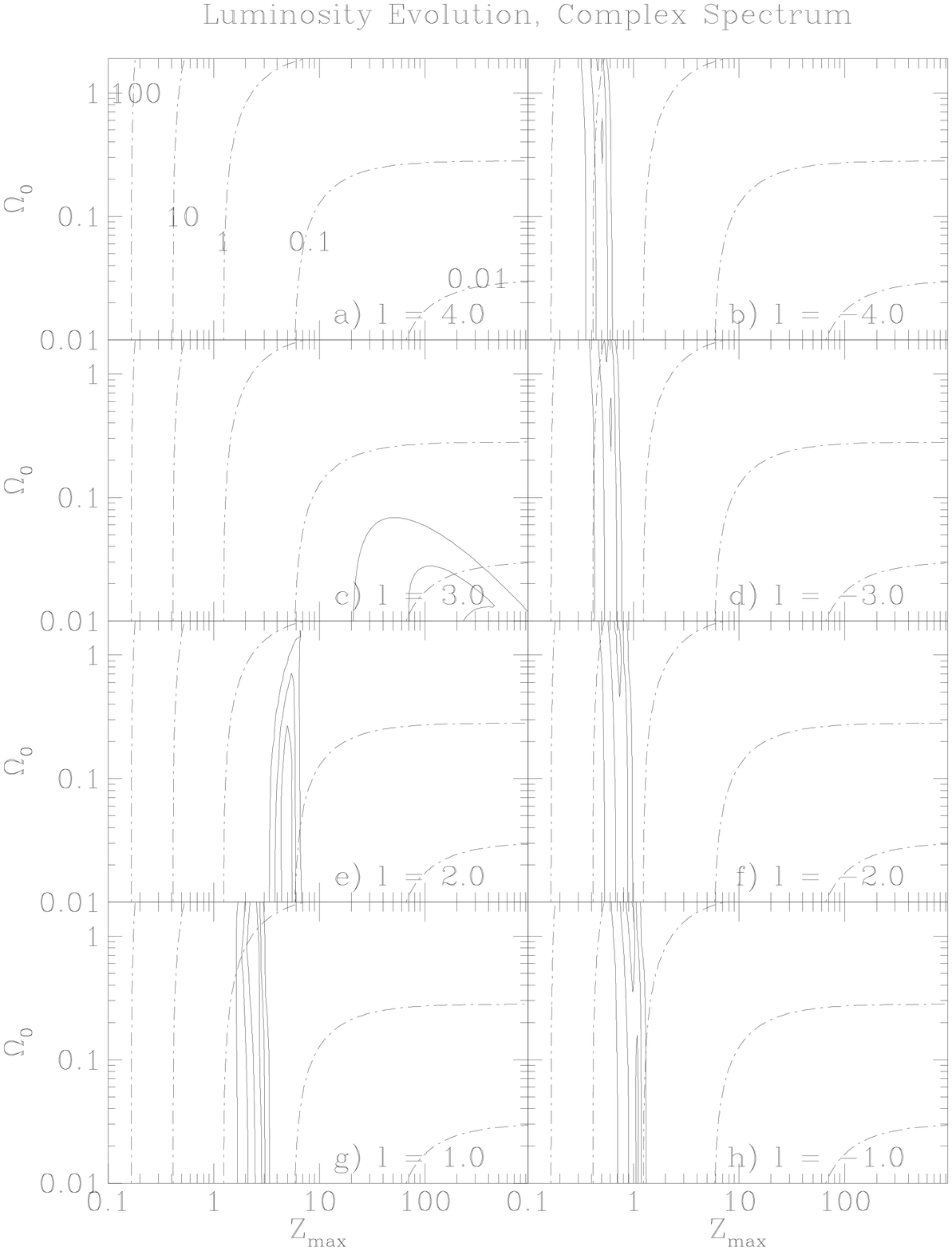 hoffset=-60 voffset=-66}{14.7cm}{21.5cm}
\FigNum{\ref{fig:lcmp}}
\end{figure}

\clearpage
\pagestyle{empty}
\begin{figure}
\PSbox{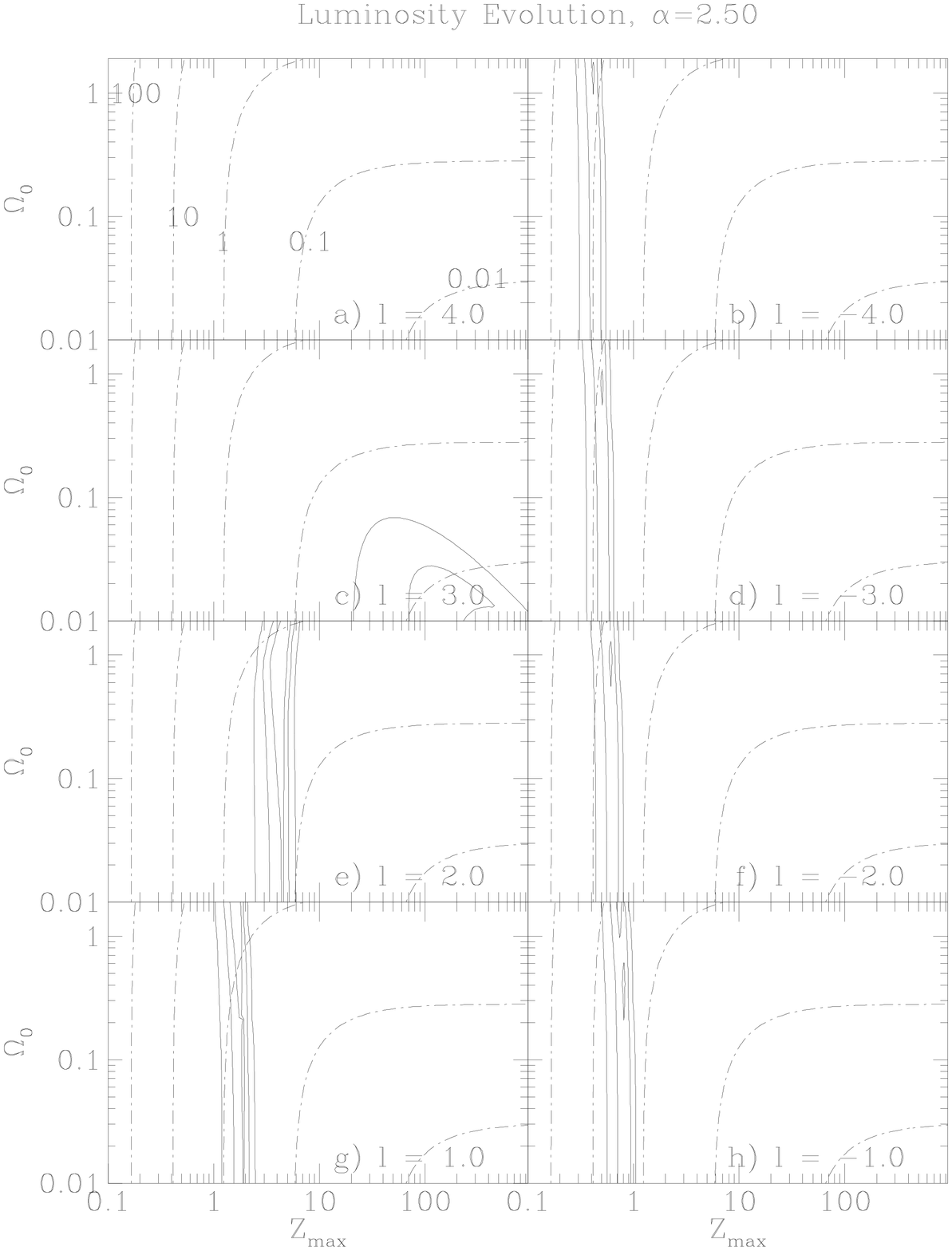 hoffset=-60 voffset=-66}{14.7cm}{21.5cm}
\FigNum{\ref{fig:l2.5}}
\end{figure}

\clearpage
\pagestyle{empty}
\begin{figure}
\PSbox{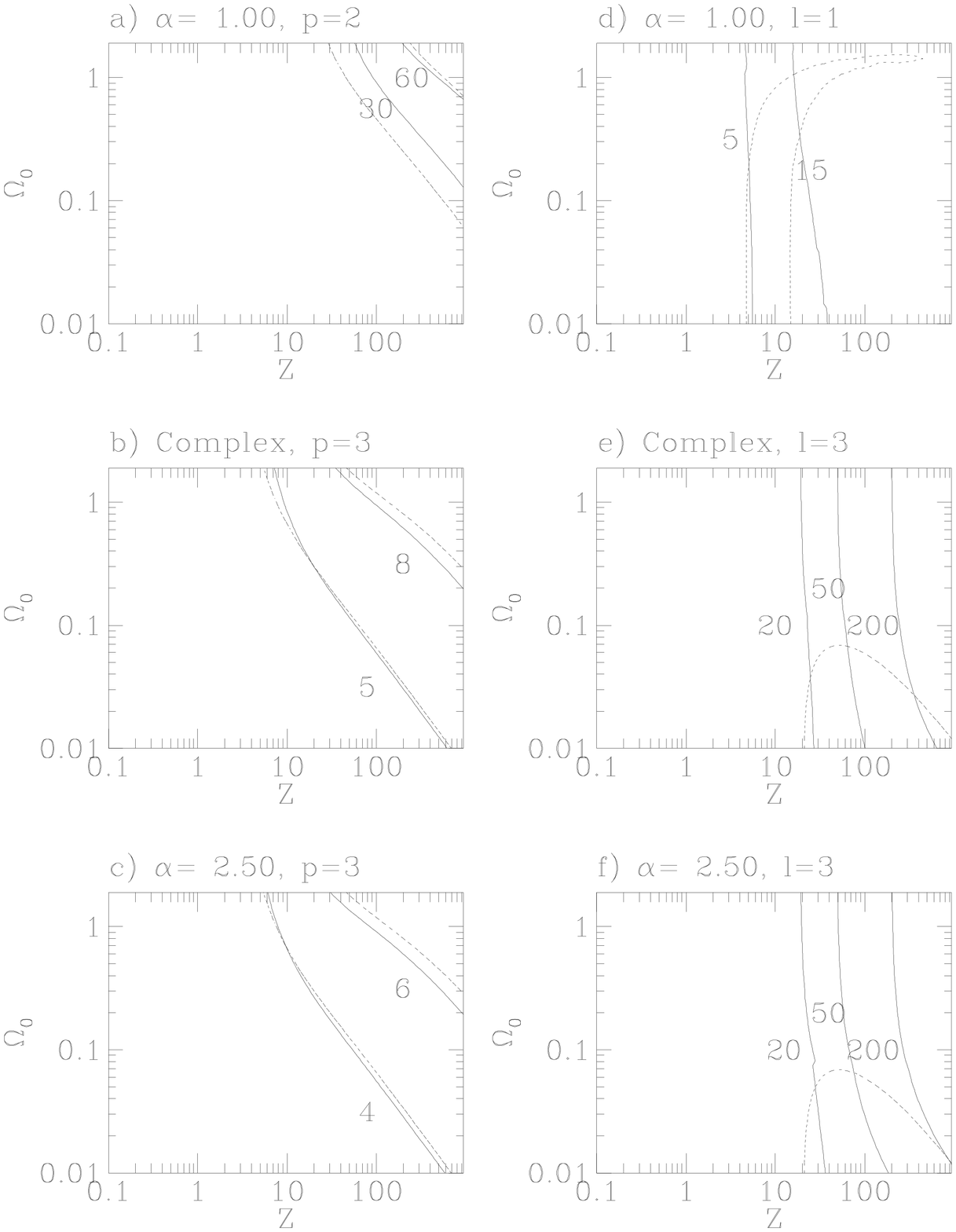 hoffset=-60 voffset=-66}{14.7cm}{21.5cm}
\FigNum{\ref{fig:Rrel2}}
\end{figure}

\clearpage
\pagestyle{empty}
\begin{figure}
\PSbox{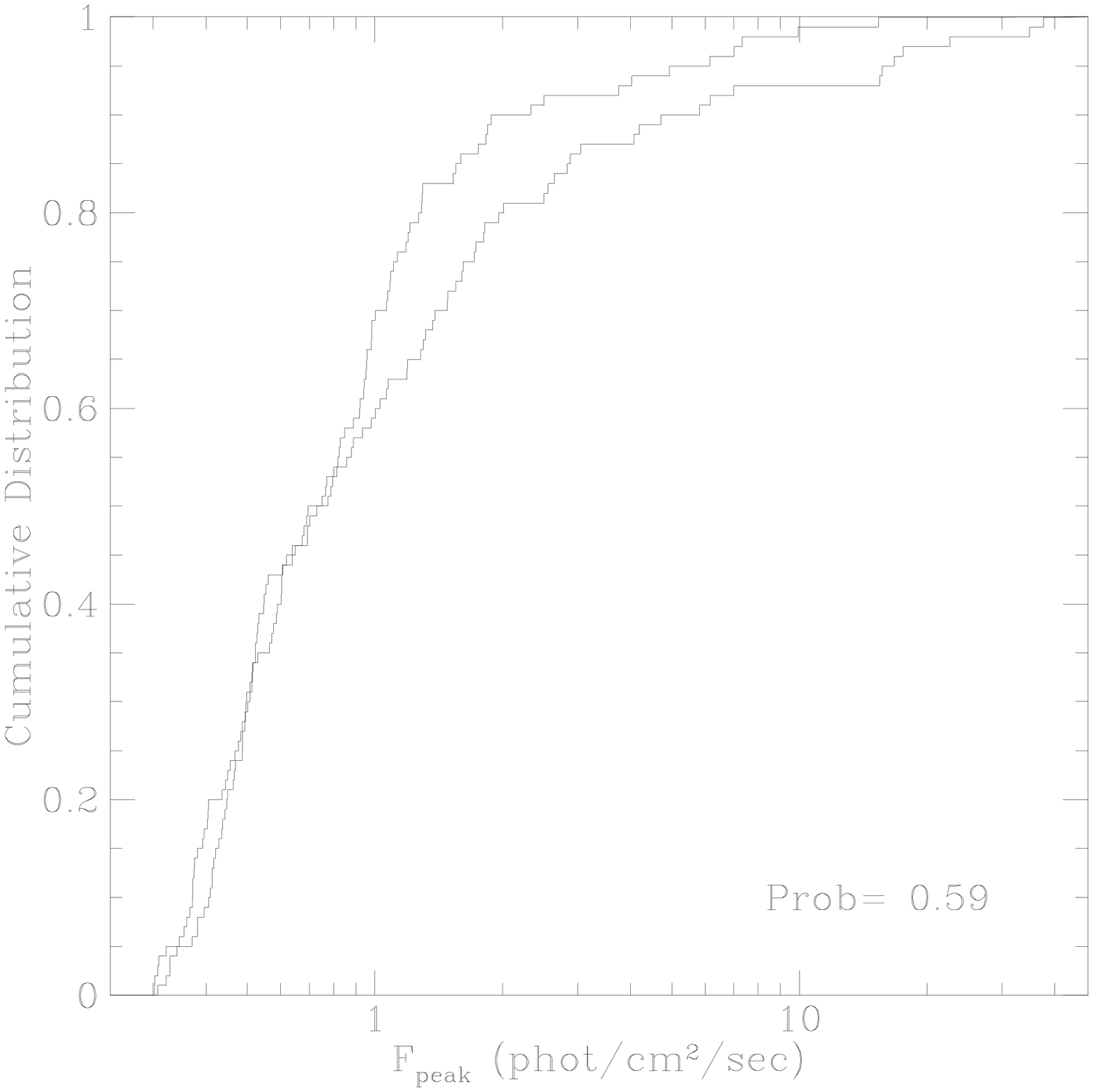 hoffset=-60 voffset=-66}{14.7cm}{21.5cm}
\FigNum{\ref{fig:reldist}}
\end{figure}

\end{document}